\documentclass[12pt, letterpaper]{article}
\usepackage[utf8]{inputenc}
\usepackage{setspace}
\usepackage{graphicx}
\usepackage{caption}
\usepackage[numbers]{natbib}
\usepackage{amsmath}

\title{Unified Finite-element Model for Transient Absorption and Raman Scattering of Vibrating Noble Metal Nanoparticles}
\author{Rachel S. Gelfand, Matthew Pelton}
\date{}

\begin{document}
\maketitle
\doublespacing

\section{Abstract}
Transient absorption and Raman scattering measurements on noble metal nanoparticles offer complimentary information on their vibrational modes and mechanical interactions with their surroundings. We have developed a comprehensive modeling tool for simulating both of these spectra based on COMSOL Multiphysics finite-element simulation software. This application can be used to predict the spectra for arbitrary geometries and metal compositions, takes into account local changes in dielectric function for the metals, and can model the small vibrational amplitudes of real transient absorption measurements. We present simulation results for gold and silver nanospheres, silver nanocubes, and gold truncated nanocubes, showing the ability to calculate relative peaks heights in Raman spectra and the ability to fit amplitudes of transient-absorption signals to experiment, and showing that Raman spectra can include contributions from modes often neglected due to symmetry considerations.

\section{Introduction}

Optical measurements of vibrating metal nanoparticles have attracted the attention of scientists for the past several decades due to the ability to provide sensitive non-contact probes of the mechanical properties of the nanoparticles and their mechanical interactions with other nanoparticles and with their surroundings \cite{Ziljstra2008}\cite{WidmerCooper2016}\cite{Voisin2002}. Strong surface plasmon resonances in the particles enable efficient excitation of acoustic vibrations and sensitive probing of these vibrations, even with sub-nm amplitudes \cite{Ahmed2017}.

The frequencies of the vibrations are dependent on the metal's material parameters, size, and geometry \cite{Crut20152}. Because of this, measurements probing the vibrational motion are able to probe the mechanical properties of nanoscale-sized objects\cite{DelFatti1999}\cite{Pelton2009}\cite{Hartland2006}, study the effects of size or geometry dispersion in ensemble samples \cite{Sauceda2012}\cite{Kelly2002}\cite{Hu2005}, and determine the compositions of multi-metallic nanoparticles \cite{Calvo2012} \cite{Yu2011}. As many of the measured samples are suspended in liquid solution, embedded in solid matrices, or placed on a solid substrate, the vibrational modes and their damping similarly report information about the nanoparticle surroundings. This has been used to probe nanoscale fluid dynamics \cite{Uthe2022} and enable sensing of environmental conditions\cite{Marty2011}.

To measure the vibrational dynamics of metal nanoparticles, two main experimental techniques are used: low-frequency Raman spectroscopy (also referred to as acoustic Raman spectroscopy) and transient absorption spectroscopy. The signals for each of these methods are dominated by a different set of vibrational modes and thus provide complementary information about vibrational dynamics \cite{Ikezawa2001} \cite{Ahmed2017}.

Transient absorption spectroscopy is a pulsed-laser method in which the sample is excited by an incident pump laser pulse and interrogated by a time-delayed probe pulse. Absorption of the pump pulse by a metal nanoparticle causes thermal expansion, inducing vibrational motion dictated by the geometry and material composition of the nanoparticle \cite{Ahmadi1996}\cite{Lamb1882}. Over the course of the mechanical vibrations, the plasmon resonance frequency and amplitude are shifted due to changes in the nanoparticle geometry, electron density distribution, and interband transition energies \cite{Ahmed2017}. The modulation of the absorption spectrum throughout the nanoparticle's vibration enables a time-resolved measurement of the particle's immediate state by the probe pulse. Analysis of transient absorption measurements has been used to return initial excitation timescales, frequencies of induced vibrational modes, and vibrational damping rates \cite{Pelton2011}. This method has been used extensively to study the vibrational frequencies of noble metal nanoparticles and their dependence on size, shape, and material constants \cite{Hartland2006} \cite{Hartland2011} \cite{Nisoli1997} \cite{Juste2005} \cite{Chen2012}. 

Low-frequency Raman spectroscopy involves the inelastic scattering of laser light off of vibrational modes in nanoparticles. (This is distinct from surface-enhanced Raman spectroscopy, in which nanoparticles are attached onto molecules or surfaces to enhance weak Raman signals \cite{Han2021}). In most cases, the vibrational frequencies are the focus of analysis due to their dependence on nanoparticle size, shape, and composition \cite{Bachelier2004}. Low-frequency Raman scattering experiments have studied elasticity at the nanoscale \cite{Hodak1999}\cite{Crut2014}\cite{Girard2017}\cite{Saviot2012}, probed the relative strengths of different scattering processes \cite{Margueriat2006}, and determined the relative probability of each vibrational mode excitation \cite{Nelet2004}. The impact of environmental conditions has been studied through observations of mechanical coupling to the surrounding media \cite{Girard2016} and other nearby nanoparticles \cite{Girard2018}\cite{Jais2011}.

In both transient absorption and low-frequency Raman scattering experiments, results are often compared to expected vibrational frequencies and damping rates. These quantities can be determined from mechanical models of the nanoparticles, but the mechanical models alone do not provide information about how the vibrations produce the measured optical signals.
They thus cannot provide information about vibrational amplitudes, because the amplitude of the optical signal does not simply correspond to the mechanical amplitude of the vibrational mode \cite{Ahmed2017}. The models are thus also of limited value for explaining apparent observations of nonlinear effects in optical measurements of nanoparticle vibrations \cite{Ahmed2022} \cite{Xiang2016}. To overcome these limitations, and to provide insight into the underlying physical mechanisms producing the measured signals, models are required that relate nanoparticle vibrations to transient absorption and Raman spectra. For highly symmetric cases, analytical models can be used \cite{Hodak1999}. Even in these cases, approximations are often used, particularly the neglect of modes that are nominally Raman inactive based on symmetry considerations. For more complex geometries, numerical methods must be used. 

In previous work \cite{Ahmed2017}, transient absorption spectra of noble metal nanoparticles were simulated using a combination of finite-element and finite-difference time-domain software packages. Because the process used two separate software packages with different geometry meshing, unrealistically large vibrational amplitudes were needed to interface the programs. Moreover, changes in dielectric function were modeled only by a single averaged value over the nanoparticle domain. In reality, the dielectric function varies over the volume during the course of its mechanical oscillation.

Saison-Francioso et al. \cite{Saison2020} expanded on this work by using finite-element software packages for both the mechanical deformation and the electromagnetic response. This enabled the modeling of local dielectric changes throughout the particle volume, which in turn enabled a deeper study into each of the contributing parameters of shape, electron density, and interband transition energy modulations. Although this method more realistically modeled transient absorption results, it required the use of three separate finite-element programs for the mechanical deformation, electromagnetic response, and data analysis.

Raman spectra have also been modeled using finite-element eigenmode solvers \cite{Wang2022} and using electromagnetic simulations based on the boundary element method \cite{Large2009}. Recent theoretical developments have provided insight into the origins of Raman signals and have opened new computational pathways \cite{Priede2022}, but have not yet been applied to more complex geometries.

The simulation method in this paper expands on these previous works by including both low-frequency Raman and transient absorption spectra calculations within a single application. In avoiding multiple simulation packages, this method does not require any complicated tricks for transferring data and eliminates much of the user input needed in the previous works. The mechanical and optical properties are simulated in the same way for both experimental methods, including the consideration of local dielectric modulation in low-frequency Raman spectroscopy simulations. Our method also takes into account all vibrational modes over the user-specified frequency range, including those traditionally considered to be Raman inactive. In addition, the tool is able to simulate realistic vibrational amplitudes matching those expected in experiments, eliminating the need for extrapolation to small deformations.

\section{Methods}

\begin{figure}
    \centering
    \includegraphics[width=0.5\linewidth]{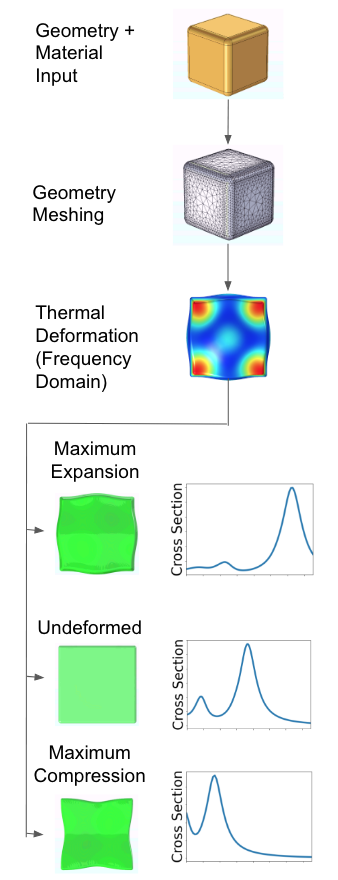}
    \caption{Flowchart for the major steps in the transient absorption calculation.}
    \label{fig:flowchartTA}
\end{figure}

\subsection{Transient Absorption Spectra}
A flowchart showing the major steps in the calculation of transient absorption spectra is shown in Fig. 1. The vibrational motion is modeled using COMSOL's Solid Mechanics module and the electromagnetic calculations are performed using COMSOL's Electromagnetic Wave, Frequency Domain solver.

The simulations begin with a calculation of the frequencies and corresponding shapes of the vibrational modes. The nanoparticle is assumed to be a linear elastic material such that the displacement of any point within the solid from its initial location can be represented by

\begin{equation}
    \nabla \cdot \mathbf{s} = \rho \mathbf{\ddot{u}}
\end{equation}

\noindent where \(\mathbf{s}\) is the 3x3 stress tensor, \(\rho\) is the density, and \(\mathbf{u}\) is the displacement vector. In the absence of initial stresses before excitation, the elastic strain and stress tensors are related by \cite{COMSOLsm}

\begin{equation}
    \mathbf{s}=\mathbf{C}:\mathbf{\epsilon}
\end{equation}

\noindent where \(\mathbf{\epsilon}\) is the 3x3 strain tensor, \(\mathbf{C}\) is the fourth-order elasticity tensor, and \(:\) denotes the double inner product operation. Assuming an isotropic, linearly elastic material, \(\mathbf{C}\) can be reduced to a tensor populated by two independent material constants. In our calculations, the material's Young's modulus and Poisson's ratio are chosen to populate this tensor (see Table 1). 

The initial condition is set by expanding the particle thermally, mimicking the expansion caused by the absorption of the pump laser pulse. This initial thermal expansion induces stress throughout the volume, which is specified by the secant coefficient of thermal expansion. The resulting equation for the total strain at a point within the volume is \cite{COMSOLsm}

\begin{equation}
    \epsilon=\alpha (T-T_{ref})
\end{equation}

\noindent where \(\alpha\) is the coefficient of thermal expansion (see Table 1), \(T\) is the temperature of the nanoparticle after laser-induced heating, and \(T_{ref}\) is the unexcited reference temperature. For the results presented in this paper, \(T_{ref}\) was set to room temperature at 293.15K.

Eqs. 1-3 are solved for the displacement, \(\mathbf{u}\), in the frequency domain. For a given frequency, the maximum displacements at each point are compiled into a displacement field. This field corresponds to the deformation of the nanoparticle at its maximum expansion in its vibration. Multiplying this field by -1 results in the deformation at the nanoparticle's maximum compression.

To model the optical response of the nanoparticle, the finite element method is used to solve Maxwell's equations in and around the nanoparticle volume as a response to an incident electromagnetic wave. The full spectrum is created by sweeping through the frequency of the incident wave. The dielectric function for the original, undeformed geometry is modeled using the Lorentz-Drude equation:

\begin{equation}
    \epsilon(\omega)=\epsilon_\infty+\sum_{m=0}^\infty\frac{f_m\omega_p}{\omega_{0,m}^2-\omega^2-i\omega\Gamma_m}
\end{equation}

\noindent where \(\epsilon_\infty\) is the permittivity at infinite frequency, \(f_m\) is a measure of the strength of each corresponding term, \(\omega_p\) is the bulk plasma frequency, \(\omega_{0,m}\) is the interband transition frequency, and \(\Gamma_m\) is the damping rate (see Table 1).

The optical response is similarly calculated for the deformed particle according to the maximally expanded and compressed deformation field found by the Solid Mechanics module. While under deformation, the dielectric function is modulated due to changes in the geometry, electron density distribution, and interband transition energies \cite{Ahmed2017}. 

\begin{table}
    \centering
    \caption{Material constants for silver and gold used in the calculations.}
    \begin{tabular}{|c|c|c|} \hline 
         Parameter & Gold & Silver\\ \hline
         \(\omega_p\) & 1.3715e16(rad/s)\(^a\)&1.368e16(rad/s)\(^b\)\\ \hline
         \(\Gamma_0\)&5.979e13(rad/s)\(^a\)&7.969e13(rad/s)\(^b\)\\ \hline
         \(\Gamma_1\)&1.07e15(rad/s)\(^a\)&0.7e15(rad/s)\(^b\)\\ \hline
         \(\Gamma_2\)&1.69e15(rad/s)\(^a\)&0.9e15(rad/s)\(^b\)\\ \hline
         \(f_0\)&1.039\(^a\)&1.029\(^b\)\\ \hline
         \(f_1\)&0.1108\(^a\)&0.06\(^b\)\\ \hline
         \(f_2\)&0.3202\(^a\)&0.165\(^b\)\\ \hline
         \(\omega_{0,1}\)&4.428e15(rad/s)\(^a\)&6.538e15(rad/s)\(^b\)\\ \hline
         \(\omega_{0,2}\)&5.746e15(rad/s)\(^a\)&7.328e15(rad/s)\(^b\)\\ \hline
         \(\epsilon_\infty\)&5.5\(^a\)&2.83\(^b\)\\ \hline
         \(\xi_1\)&-0.8\(^c\)&-0.3\(^d\)\\ \hline
         \(\xi_2\)&-9\(^c\)&-9\(^d\)\\ \hline
         Young's Modulus&7e10(Pa)\(^e\)&8.3e10(Pa)\(^e\)\\ \hline
         Poisson's Ratio&0.44\(^e\)&0.37\(^e\)\\ \hline
         \(\rho\)&1.93e4(kg/m\(^3\))\(^e\)&1.05e4(kg/m\(^3\))\(^e\)\\ \hline
         \(\alpha\)&1.426e-7(1/K)\(^e\)&1.89e-7(1/K)\(^e\)\\ \hline
    \end{tabular}
    \\,
    \caption*{\(^a\) Taken from \cite{Ahmed2017}, found by fitting Eq. 4 to experimental data of \cite{Olmon2012}
    \\ \(^b\) Taken from \cite{Ahmed2017}, found by fitting Eq. 4 to experimental data of \cite{Yang2015}
    \\ \(^c\) Retrieved from \cite{Szczepanek1974}
    \\ \(^d\) Retrieved from \cite{Tups1984}
    \\ \(^e\) From COMSOL's material library \cite{COMSOL}
    }
    \label{tab:my_label}
\end{table}

The difference in the free electron density, \(N\), affects the dielectric function throughout the material through the bulk plasmon frequency, \(\omega_p\), according to the following relation:

\begin{equation}
    \omega_p=\sqrt{\frac{Ne^2}{\epsilon_0m}}
\end{equation}

\noindent where \(e\) is the electron charge, \(\epsilon_0\) is the permittivity of free space, and \(m\) is the electron mass. The bulk plasmon frequency for the deformed particle, denoted by \(\omega_{p}'\), can be calculated from the undeformed frequency along with the volumetric strain \(dV/V\):

\begin{equation}
    \omega_{p}'=\omega_p\sqrt{\frac{1}{1+\frac{dV}{V}}}
\end{equation}

\noindent The differential volumetric strain is calculated point-by-point throughout the nanoparticle volume using the deformation field \(\mathbf{u}\) found by the Solid Mechanics module. The interband transition energies are similarly modulated through the volumetric strain and the deformation potential, \(\xi_m\) for each associated band (see Table 1):

\begin{equation}
    \omega_{0,m}'=\omega_{0,m} + \frac{\xi_mdV}{\hbar V}
\end{equation}

The important outputs of the electromagnetic portion of the calculation are the electric field \(\mathbf{E}\), the magnetic field \(\mathbf{H}\), and the current density \(\mathbf{J}\). The electric and magnetic fields are separated into incident and scattered field components: 

\begin{align}
    \mathbf{E}_{tot} &= \mathbf{E}_{inc} + \mathbf{E}_{sc} \\
    \mathbf{H}_{tot} &= \mathbf{H}_{inc} + \mathbf{H}_{sc}
\end{align}

The calculated fields are used to calculate the absorption and scattering cross sections for the undeformed, maximally expanded, and maximally compressed geometries. The absorption cross-section for a nanoparticle is calculated as the power dissipated throughout the nanoparticle volume V \cite{COMSOLewfd}:

\begin{equation}
    \sigma_{abs}=Q_{loss}=\int_V \mathbf{J}\cdot\mathbf{E}_{sc} dV
\end{equation}

The scattering cross-section is determined by integrating the scattered Poynting vector \(\mathbf{S}_{sc}=\mathbf{E}_{sc} \times \mathbf{H}_{sc}\), over a sphere \(A\) centered on and entirely containing the nanoparticle:

\begin{equation}
    \sigma_{sc}=- \frac{1}{S_{inc}}\int_A \mathbf{S}_{sc} \cdot \hat{\mathbf{e}}_r dA
\end{equation}

\noindent where \(S_{inc}\) is the magnitude of the incident Poynting vector, and \(\hat{\mathbf{e}_r}\) is the vector normal to the integration sphere \cite{Stamnes2015}.

The extinction cross-section in each of the three geometries is the sum of the absorption and scattering cross sections:

\begin{equation}
    \sigma_{ext}=\sigma_{sc}+\sigma_{abs}
\end{equation}

For the undeformed, maximally compressed, and maximally expanded geometries, each spectrum is calculated as a function of incident wavelength. The difference between the extinction spectra for the deformed and undeformed geometries corresponds to the transient spectrum typically measured experimentally.

\subsection{Raman Scattering Spectra}

\begin{figure}
    \centering
    \includegraphics[width=0.4\linewidth]{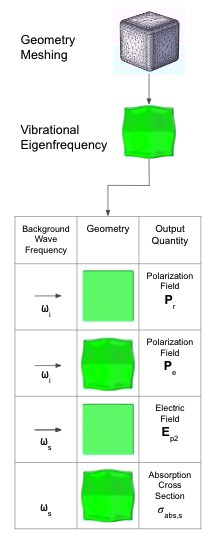}
    \caption{Flowchart detailing the major steps in the Raman scattering calculation. The rows in the table represent each of the four electromagnetic simulations and their associated output quantity used in the final amplitude equation in Eq. 17.}
    \label{fig:flowchatRaman}
\end{figure}

A flowchart for the Raman scattering calculations is shown in Fig. 2. As for the transient absorption simulations, the first step in the Raman scattering simulations is the calculation of the vibrational mode frequencies and displacement fields using COMSOL's Solid Mechanics module. The nanoparticle is similarly assumed to be a linearly elastic material with properties dictated by the density, Young's modulus, and Poisson's ratio. In this case, however, the vibrational modes are calculated using COMSOL's built-in eigenvalue solver. This returns the eigenfrequency corresponding to the vibrational mode frequency (\(\omega_{vib}\)) and eigenvector displacement field (\(\mathbf{u}\)) of each mode. 

Because of the nature of these eigenvalue solutions, the displacement field for each mode has an overall arbitrary amplitude which must be normalized.
A normalization factor, \(n_i\), can be introduced to the displacement field. \(u_i\), for mode. \(i\). such that each mode contains one quantum of energy, \(hf_i\), where \(f_i\) is the vibrational frequency of the mode. Choosing the first mode to normalize all other modes to, the normalization factor for the displacement field of the \(i\)th vibrational mode is

\begin{equation}
    n_i=\sqrt{\frac{E_{1}*f_1}{E_{i}*f_i}}
\end{equation}

\noindent where \(E_i\) is the total elastic strain energy of the mode.  For a Hookean solid, this energy is calculated from the magnitude of the displacement field through

\begin{equation}
    E_i \propto \int_V {u_i}^2 dV
\end{equation}

Raman scattering involves three processes: (1) absorption of an incoming photon with frequency \(\omega_i\) and excitation of a surface plasmon with frequency \(\omega_{p1}\), (2) decay of the surface plasmon into a vibrational state with frequency \(\omega_{vib}\) and a secondary surface plasmon with frequency \(\omega_{p2}\), and (3) decay of this second surface plasmon into a scattered photon with frequency \(\omega_s\). All of the intermediate steps occur virtually, and the overall scattering is an instantaneous process described by third-order perturbation theory. According to this theory, the relative scattering amplitude for a single input frequency in a Raman scattering experiment is proportional to \cite{Bachelier2007}

\begin{equation}
    |\sum\limits_{vib}\frac{\int \mathbf{E}_s \cdot \mathbf{P}_{p2}dV}{(\hbar\omega_s - \hbar\omega_{p2}+i\Gamma_{p2})}\int \mathbf{E}_{p2}\cdot [\mathbf{P}_e-\mathbf{P}_r]|^2
\end{equation}

\noindent where the sum is over all of the vibrational modes, \(\mathbf{E}\) and \(\mathbf{P}\) denote the electric field and polarization field, respectively, and \(\Gamma\) is the plasmon linewidth. The subscript details the step of the scattering process at which the fields are calculated, and integrals are taken over all space. \(\mathbf{P}_e\) is the polarization field for the deformed particle during vibration and \(\mathbf{P}_r\) is the polarization field for the undeformed particle, both calculated according to excitation with frequency \(\omega_i\).

This can be further simplified by recognizing that the absorption cross-section of the vibrating nanoparticle taken at \(\omega_s\) is proportional to

\begin{equation}
    \sigma_{abs,s} \propto |\frac{\int \mathbf{E}_s \cdot \mathbf{P}_{p2}dV}{(\hbar\omega_s - \hbar\omega_{p2}+i\Gamma_{p2})}|^2
\end{equation}

\noindent This absorption cross section can be calculated in the same manner as for the transient absorption spectra according to Eq. 9. The resulting Raman scattering amplitude for incoming frequency \(\omega_i\) is proportional to

\begin{equation}
    |\sum\limits_{vib}\sigma_{abs,s} [\int \mathbf{E}_{p2}\cdot [\mathbf{P}_e-\mathbf{P}_r]] dV]|^2
\end{equation}

\noindent For each vibrational mode, the solution to this equation is found using COMSOL's electromagnetic wave frequency domain solver in the same fashion as for the transient absorption calculation.

For each of the variables in Eq. 17, a different set of incident wave frequencies and deformations are needed to account for their appropriate stage in the scattering process. The electromagnetic wave simulation is therefore run four times for each of the necessary inputs to the simulation. A summary of these separate calculations and their associated input parameters can be found in the bottom half of Fig. 2.

The result of those calculations is a relative scattering amplitude, \(a_i\), for each mode, \(i\), with Raman shift, \(x_i\). In reality, there is a broadening of each Raman peak in the experimental spectra due to many factors, including the plasmon lifetime and the instrument response function. To account for this and to more closely reproduce experimental spectra, the overall spectrum is represented by a Lorentzian function for each calculated vibrational mode:

\begin{equation}
    L(x)=\sum_{i}\frac{a_i\Gamma}{(x-x_{0i})^2+(\frac{1}{2}\Gamma)^2}
\end{equation}

\noindent where the linewidth \(\Gamma\) is chosen to match experiment.

\section{Results}

In order to illustrate the capabilities of this method, we calculated transient absorption and Raman spectra for gold and silver nanoparticles of a number of different sizes and shapes.

\subsection{Nanospheres}

The simulated Raman spectrum for a silver sphere with a diameter of 5nm can be found in Fig. 3. Due to the high symmetry in this situation, the spectra can also be calculated from Lamb theory and Maxwell's equations \cite{Bachelier2004}. The corresponding frequencies derived from this theoretical calculation are shown as vertical lines in Fig. 3. Our results show general agreement with the frequencies calculated using this analytical model with differences most likely due to differences in assumed elastic constants.

The analytical method assumes a constant dielectric function throughout the particle. Together with the assumption that the deformation of the particle is negligible, this means that there are many modes for which the inner product of the electric field and the modulation in the polarization field, as calculated in Eq. 15, perfectly cancel when integrating over the nanoparticle volume. These modes are often referred to as Raman inactive and have often been ignored in theoretical models.

Recently, nano-sized materials have been shown to produce measurable Raman peaks even for ``inactive" modes \cite{Kuok2003}. At the nanoscale, the perfect cancellation breaks down as the distortion of the nanoparticle over the course of its deformation causes the breaking of perfect spherical symmetry and the contribution from Eq. 15 can no longer be neglected.
Indeed, these ``inactive" modes contribute to our calculated spectra, demonstrating that all modes should be considered in order to most accurately predict the Raman spectra.

\begin{figure}
    \centering
    \includegraphics[width=0.5\linewidth]{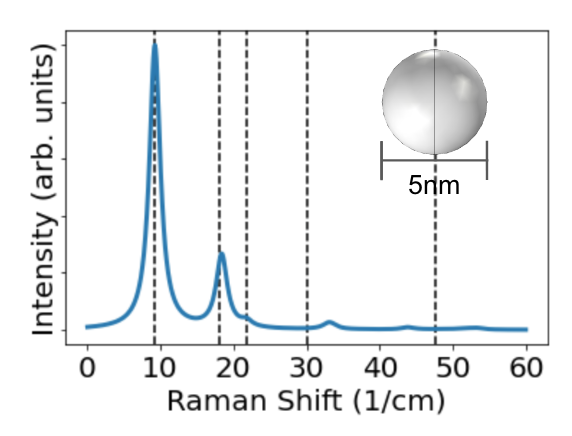}
    \caption{Simulated Raman spectra for a 5nm diameter silver sphere. Majorly contributing theoretically derived frequencies from \cite{Bachelier2004} are noted by vertical dashed lines.}
    \label{fig:Ramansilversphere}
\end{figure}

Fig. 4 shows the calculated Raman spectrum for a 3.2 nm gold sphere and comparison to experimental data reported by Mankad et al. \cite{Mankad2013}. The good agreement validates the choice of mechanical material constants used in the calculations.

\begin{figure}
    \centering
    \includegraphics[width=0.5\linewidth]{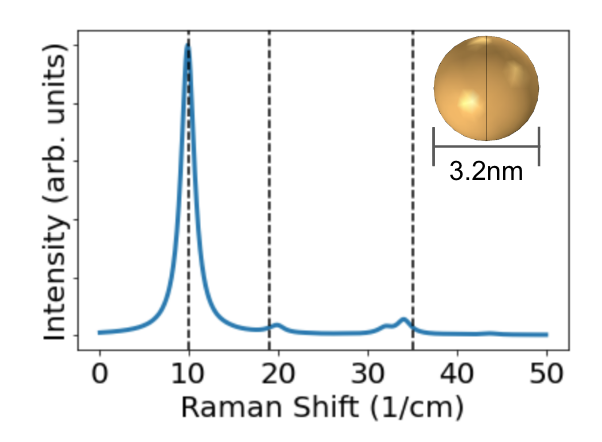}
    \caption{Simulated Raman spectrum for a 3.2nm diameter gold nanosphere. Locations of experimental peaks from \cite{Mankad2013} are noted by vertical dashed lines.}
    \label{fig:Ramangoldsphere}
\end{figure}

The effects of sample size distribution on calculated Raman spectra were explored by comparing to the experimental Raman spectrum of gold spheres with a mean diameter of 5.8 nm and standard deviation of 0.75 nm \cite{Portales2010}. We simulated the spectra of gold spheres with diameters of 5.8 nm, 6.55 nm, and 5.05 nm (corresponding to the mean diameter, one standard deviation above the mean, and one standard deviation below the mean). The 5.8 nm diameter spectra were normalized to a maximum intensity of 1, and the calculated 6.55 nm and 5.05 nm diameter spectra were normalized to a maximum intensity of exp(-1/2) to account for their relative proportions in the Gaussian size distribution. The experimental and all simulated results are shown in Fig. 5.

The experimental spectrum includes background Rayleigh scattering, which shifts the maximum of the Raman scattering peak to slightly lower frequencies, accounting for the difference in apparent peak frequencies between the experimental and calculated spectra. The range of the simulated spectra matches with the linewidth of the experimental spectrum, illustrating broadening due to the nanoparticle size dispersion.

\begin{figure}
    \centering
    \includegraphics[width=0.5\linewidth]{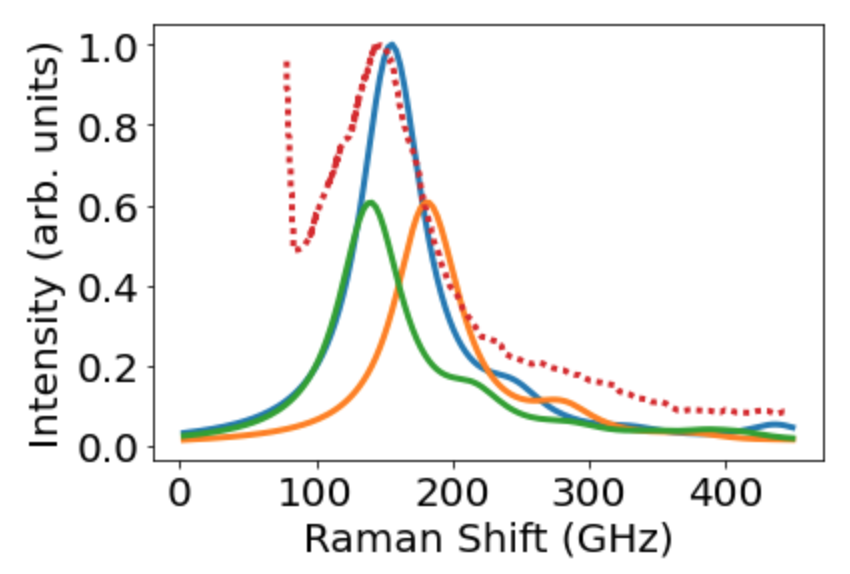}
    \caption{Simulated and experimental Raman spectra for a sample of gold spheres with a Gaussian size distribution. Experimental data (red) are reproduced from \cite{Portales2010}. Simulated spectra are shown for spheres with diameters of 5.8 nm (mean diameter, blue), 6.55 nm (one standard deviation above mean, green), and 5.05 nm (one standard deviation below mean, orange), with peak heights adjusted to correspond to the respective Gaussian weights.}
    \label{fig:SphereRamanSizeDistribution}
\end{figure}

Fig. 6 shows transient absorption calculations for silver spheres with a diameter of 60nm, and Fig. 7 shows results for gold spheres with a diameter of 50nm.
Experimental results have been reported for 50nm gold spheres in water \cite{Chakraborty2021}. In this work, the change in absorption spectra at a specified delay time was fit to the difference of Lorentzians corresponding to the plasmon resonance with and without laser excitation, and plasmon peak shifts were determined from these fits as a function of pump-probe delay time \cite{Pelton2009}. These experimental values are represented as points in Fig. 6e. An exponentially decaying signal corresponding to heating and subsequent cooling has been subtracted from the results to isolate the oscillations due to particle vibration.

For comparison to calculations, the vibrational frequency, \(\omega_{vib}\), was taken to be the lowest frequency mode shown in Fig. 6b. The initial temperature was chosen such that the calculated peak shift matches the largest peak shift measured experimentally. For a given vibrational mode, the shape of the peak shift function over time is expected to be a damped oscillation \cite{Pelton2009}\cite{Pelton2011}:

\begin{equation}\label{eq:dampedep}
    \Delta A = \Delta\omega * exp (-\frac{t}{\tau_1}) sin (\omega_{vib} t + \phi)
\end{equation}
\(\tau_1\) and \(\phi\) were chosen to best match experimental results. The corresponding simulation output is shown in Fig. 6e as a red line.

The displacement of the nanoparticle surface in the breathing mode vibration due to the fitted temperature increase is 3 pm (see Fig. 6b). This illustrates how comparison between our calculations and experimental data enables the precise determination of vibrational amplitudes.

\begin{figure}
    \centering
    \includegraphics[width=1\linewidth]{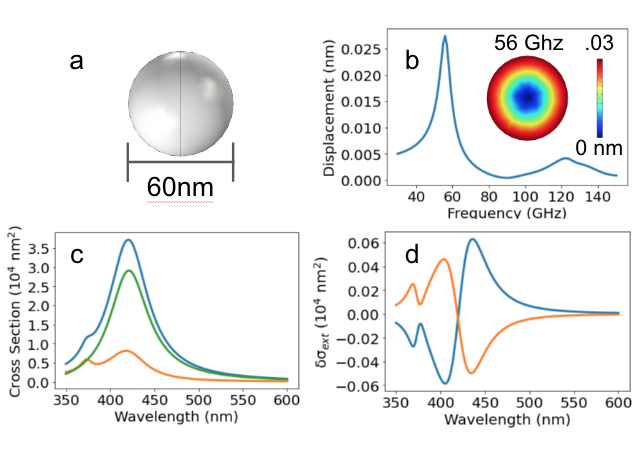}
    \caption{Simulated transient absorption spectra for a 60nm diameter silver sphere. Thermal expansion temperature was set to 300K. (a) Geometry input into the simulation. (b) Displacement of a point on the surface of the sphere as a function of frequency. Inset: cross-sectional displacement for the 56 GHz mode. (c) Extinction (blue), absorption (green), and scattering (orange) cross sections for the undeformed geometry as a function of incident wavelength. (d) Change in extinction cross section vs incident wavelength for maximum expansion (blue) and contraction (orange) for the 56 GHz vibrational mode.}
    \label{fig:TAsilversphere}
\end{figure}

\begin{figure}
    \centering
    \includegraphics[width=0.3\linewidth]{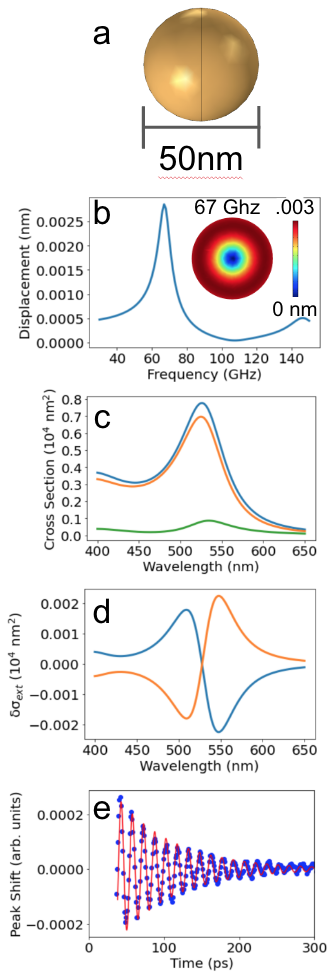}
    \caption{Simulated transient absorption spectra and kinetics for a 50nm diameter gold sphere. Thermal expansion temperature was set to 294K. (a) Geometry input into the simulation. (b) Displacement of a point on the surface of the sphere as a function of frequency. Inset: cross-sectional displacement for the 67 GHz mode. (c) Extinction (blue), absorption (green), and scattering (orange) cross sections for the undeformed geometry as a function of incident wavelength. (d) Change in extinction cross section vs incident wavelength for the maximum expansion (blue) and compression (orange) for the 67 GHz vibrational mode. (e) Experimental transient absorption results from \cite{Chakraborty2021} (blue) and the corresponding simulation output for peak shift over time. Exponential damping is assumed to account for mechanical energy losses in the nanoparticle and its surroundings.}
    \label{fig:TAgoldsphere}
\end{figure}

\subsection{Nanocubes}

Another common geometry for metal nanoparticles is the cube. Raman spectroscopy on this geometry has not yet been reported. A simulated Raman spectrum for a silver cube with a side length of 5nm can be found in Fig. 8. The edges of the cube have all been rounded to more accurately reproduce the geometry in real samples.

\begin{figure}
    \centering
    \includegraphics[width=0.5\linewidth]{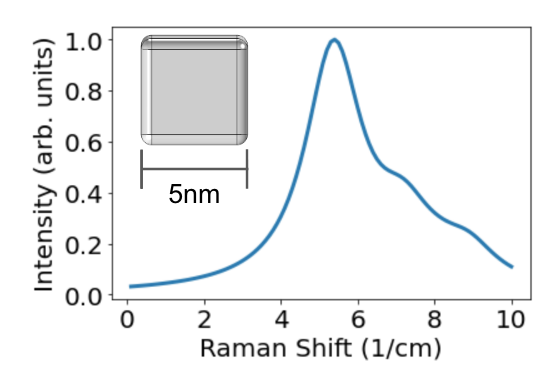}
    \caption{Simulation results for a silver cube with a side length of 5nm. Edges have been rounded to model realistic particle shapes.}
    \label{fig:Ramansilvercubel}
\end{figure}

Calculated transient absorption spectra for a silver cube with length of 35nm can be found in Fig. 9. Petrova et. al. reported an experimental vibrational frequency of 52.6 GHz for silver cubes with side lengths of 35.5 +/- 3.4 nm \cite{Petrova2007}. Our major contributing mode was calculated to have a frequency of 60 GHz. Taking into account significant error in the reported side lengths, our simulation output is consistent with the measured frequencies.

\begin{figure}
    \centering
    \includegraphics[width=1\linewidth]{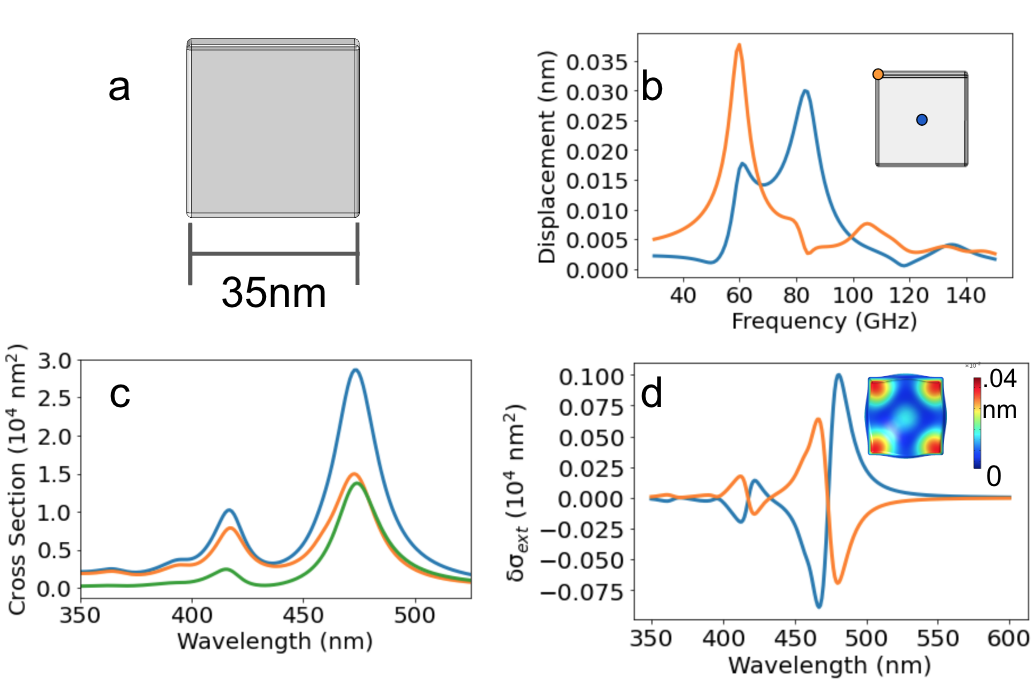}
    \caption{Simulated transient absorption spectra for a silver cube with a side length of 35 nm. Thermal expansion temperature was set to 300 K, and cube edges were rounded to model realistic particle shapes. (a) Geometry input into the simulation. (b) Displacement of two points on the surface of the cube as a function of frequency. Inset: location of each point with colors matching the corresponding spectra. (c) Extinction (blue), absorption (green), and scattering (orange) cross sections for the undeformed geometry as a function of incident wavelength. (d) Change in extinction cross section vs incident wavelength for maximum expansion (blue) and contraction (orange) for the 60 GHz vibrational mode. Inset shows the displacement for the 60 GHz mode along a cross-section taken through the center of the cube.}
    \label{fig:TAsilvercube}
\end{figure}

Previously, we used our method to simulate silver nanocubes with an octahedral gold core \cite{Ahmed2022}. The model predicted a nonlinear dependence of the optical signal on the vibrational amplitude of one of the modes. This nonlinearity may account for apparent frequency mixing in the measured signal and illustrates the ability of our simulations to more deeply probe the contributing causes of experimental transient absorption signals.

\subsection{Truncated Nanocubes}

Truncated nanocubes are often synthesized \cite{Wu2010}, but their transient absorption and Raman spectra have not previously been calculated. The transient absorption spectra for a gold truncated nanocube with side length of 35nm can be found in Fig. 10. This demonstrates the simulation's ability to predict transient absorption spectra for nanoparticles with more complex shapes.

\begin{figure}
    \centering
    \includegraphics[width=0.445\linewidth]{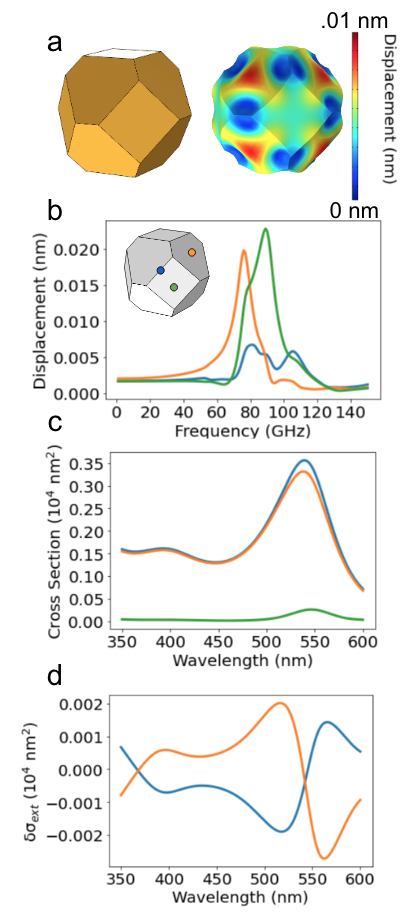}
    \caption{Simulated transient absorption spectra for a truncated gold cube with a side length of 35nm. Thermal expansion temperature was set to 300 K. (a) Geometry input into the simulation (left) and the displacement field for the first vibrational mode (right). (b) Displacement of three points on the surface of the particle as a function of frequency. Inset: location of each point with colors matching the corresponding spectra. (c) Extinction (blue), absorption (green), and scattering (orange) cross sections for the undeformed geometry as a function of incident wavelength. (d) Change in extinction cross section vs incident wavelength for maximum expansion (blue) and contraction (orange) for the first vibrational mode.}
    \label{fig:TAtruncated}
\end{figure}

\section{Conclusion}
We have developed a simulation method and associated package for simulating both the transient absorption and Raman scattering spectra for vibrating noble metal nanoparticles. This method improves on previous models and analytical formulations due to its inclusion of point-by-point modulation of the dielectric function, realistic vibrational amplitudes, and treatment of both measurement methods within a single software interface. The framework holds for arbitrary geometry and metal composition, providing a single tool for a wide range of different applications.

Because the method does not make any symmetry-based arguments for neglecting certain vibrational modes, it predicts contributions of traditionally inactive modes to the Raman spectra for spherical particles. In addition, our method enables the extraction of vibrational amplitudes from transient absorption data and thus the consideration of more complex phenomena such as nonlinear dependence of optical signal on vibrational amplitude.

Further extension of this work can be carried out by including nanoparticle environment effects in the simulation package. Additional complexity can similarly be added to more completely model additional physics including material anisotropy, temperature-induced changes in material properties, and radiation pressure.

\section{Acknowledgements}

We acknowledge funding from the US National Science Foundation under grant number DMR-1554895.
We would like to thank Nicolas Large, Aftab Ahmed, and Jeffrey Guest for helpful discussions.

\singlespacing
\bibliography{Sources}
\bibliographystyle{unsrtnat}
\begin{figure}
    \centering
    \title{For Table of Contents Only}
    \includegraphics[width=0.9\linewidth]{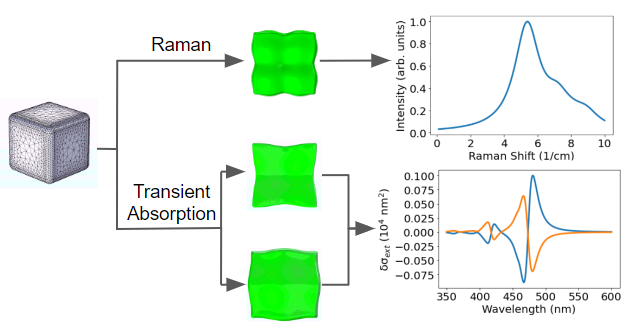}

\end{figure}
\end{document}